\documentclass[12pt]{article}
\usepackage{graphicx,latexsym}
\usepackage[noadjust]{cite}
\usepackage{amsmath} 
\usepackage{amssymb}  
\usepackage{url}
\def\proof{{\em Proof: }}

\newtheorem{theorem}{Theorem}
\newtheorem{corollary}{Corollary}

\newtheorem{lemma}{Lemma}

\newtheorem{definition}{Definition}
\def\eop{\hfill$\Box$}
\def\C{{\mathbb C}}

\title{Multipartite separability of Laplacian matrices of graphs}

\author{Chai Wah Wu\\
\small IBM T. J. Watson Research Center\\[-0.8ex]
\small P. O. Box 704, Yorktown Heights, NY 10598, USA\\[-0.8ex]
\small \texttt{cwwu@us.ibm.com}}

\date{September 4, 2008\\
\small Mathematics Subject Classifications: 81P15, 81P68}

\begin{document}
\maketitle

\begin{abstract}
Recently, Braunstein et al. \cite{braunstein:laplacian:2006} introduced normalized Laplacian matrices of graphs as density matrices in quantum mechanics and studied the relationships between quantum physical properties and graph theoretical properties of the underlying graphs.  We provide further results on the multipartite separability of Laplacian matrices of graphs.  In particular, we identify complete bipartite graphs whose normalized Laplacian matrix is multipartite entangled under any vertex labeling.  Furthermore, we give conditions on the vertex degrees such that there is a vertex labeling under which the normalized Laplacian matrix is entangled.  These results address an open question raised in \cite{braunstein:laplacian:2006}.
Finally, we extend some of the results in \cite{braunstein:laplacian:2006,wang:tripartite:2007} to the multipartite case and show that the Laplacian matrix of any product of graphs (strong, Cartesian, tensor, categorical, etc.) is multipartite separable.  
\end{abstract}

\section{Introduction}
The object of study in this paper are density matrices of quantum mechanics.  Density matrices are used to describe the state of a discrete quantum system and are fundamental mathematical constructs in quantum mechanics.  They play a key role in the design and analysis of quantum computing and information systems \cite{nielsen-quantum-2002}.

\begin{definition}
A complex matrix $A$ is a {\em density matrix} if it is Hermitian, positive semidefinite and has unit trace.
\end{definition}

{\em Remark:} In this paper we will often use the following simple fact: $\frac{1}{tr(A)} A$ is a density matrix if $A$ is Hermitian, positive semidefinite and has a strictly positive trace.  We will refer to $\frac{1}{tr(A)}A$ as a {\em normalization} of $A$.

\begin{definition}
A complex matrix $A$ is row diagonally dominant if $A_{ii} \geq \sum_{j\neq i} |A_{ij}|$ for all $i$.
\end{definition}

By Gershgorin's circle criterion, all the eigenvalues of a row diagonally dominant matrix has nonnegative real parts.  Thus a nonzero Hermitian row diagonally dominant matrix is positive semidefinite and has a strictly positive trace, and such a matrix normalized is a density matrix.

A key property of a density matrix is its separability.  The property of nonseparability plays an important role in generating the myriad of counterintuitive phenomena in quantum mechanics.

\begin{definition}
A density matrix $A$ is {\em separable} in $\C^{p_1}\times \C^{p_2} \times \cdots \times \C^{p_m}$
if it can be written as $A = \sum_i c_i A_i^1 \otimes \cdots \otimes A_i^m$ where
$c_i\geq 0$, $\sum_i c_1 = 1$ and $A_i^j$ are density matrices in $\C^{p_j\times p_j}$.
A density matrix is {\em entangled} if it is not separable.
\end{definition}

\section{Laplacian matrices as density matrices}
The Laplacian matrix of a graph is defined as
$L = D-A$ , where $D$ is the diagonal matrix of the vertex degrees and $A$ is the adjacency matrix.  The matrix $L$ is symmetric and row diagonally dominant, and therefore for a nontrivial\footnote{A graph is {\em trivial} if it has no edges.  In this case the Laplacian matrix is the zero matrix and has zero trace.} graph the matrix
$\frac{1}{tr L}L$ is a density matrix.
In Ref. \cite{braunstein:laplacian:2006}, such normalized Laplacian matrices are studied as density matrices and quantum mechanical properties such as entanglement of various types of graph Laplacian matrices are studied.
This approach was further investigated by \cite{wu:separable:2006} who showed that the Peres-Horodecki necessary condition for separability is equivalent to a condition on the partial transpose graph, and that this condition is also sufficient for separability of block tridiagonal Laplacian matrices and Laplacian matrices in $\C^2\times \C^q$.   In \cite{wang:tripartite:2007} the tripartite separability of normalized Laplacian matrices is studied.  In \cite{braunstein:laplacian_graph:2006} several classes of graphs were identified whose separability are easily determined.

As noted in \cite{braunstein:laplacian:2006}, the separability of a normalized Laplacian matrix of a graph depends on the labeling of the vertices.  In the sequel, unless otherwise noted (for example, in Theorem \ref{thm:bipartite1}), we will assume a specific Laplacian matrix (and thus a specific vertex labeling) when we discuss separability of Laplacian matrices of graphs.
A vertex labeling can be defined as:

\begin{definition}
For $n= p_1p_2\cdots p_m$, a {\em vertex labeling} is a bijection between
$\{1,\dots , n\}$ and $\{1,\dots , p_1\}\times \{1,\dots , p_2\} \times \cdots \times
\{1,\dots ,p_m\}$.
\end{definition}

\section{Conditions for multipartite entanglement} \label{sec:bipartite}
In this section, we consider unweighted graphs, i.e. the adjacency matrix is a $0$-$1$ matrix.

\begin{definition}
Given a graph $\cal G$ with vertices  $V\times W$, the partial transpose graph ${\cal G}^{pT}$ is a graph with vertices
$V\times W$ and edges defined by:

$\{(u,v),(w,y)\}$ is an edge of $\cal G$ if and only if $\{(u,y),(w,v)\}$ is an edge of ${\cal G}^{pT}$
\end{definition}
Note that the partial transpose graph depends on the specific labeling of the vertices.  The partial transpose graph is useful in determining separability of the Laplacian matrix of a graph with the same vertex labeling.
In \cite{braunstein:laplacian_graph:2006,wu:separable:2006}
the following necessary condition for separability is shown:

\begin{theorem}
If the normalized Laplacian matrix of $\cal G$ is separable then each vertex of $\cal G$ has the same degree as the same vertex of ${\cal G}^{pT}$. \label{thm:degree}
\end{theorem}

\begin{corollary}
If the normalized Laplacian matrix of $\cal G$ is separable then each vertex of $\overline{\cal G}$ has the same degree as the same vertex of ${\overline{\cal G}}^{pT}$. \label{cor:degree-comp}
\end{corollary}
\proof Follows from the fact that the degree condition in Theorem \ref{thm:degree} is true for a graph ${\cal G}$ if and only if it is true for the complement graph ${\cal G}$.\eop

In \cite{wu:separable:2006} the following sufficient condition for separability is shown:

\begin{theorem}
For a graph ${\cal G}$, if for all $1\leq i,j\leq p_1$, $1\leq k\leq p_2$, $i\neq j$, the number of edges from vertex $(v_i,w_k)$ to vertices of the form $(v_j,\cdot )$ is the same as the number of edges from vertex $(v_j,w_k)$ to
vertices of the form $(v_i,\cdot )$, then the normalized Laplacian matrix of ${\cal G}$ is separable in 
$\C^{p_1}\times \C^{p_2}$.
\end{theorem}

\subsection{Complete bipartite graphs}
\begin{theorem}
Let $n=p_1p_2\cdots p_m$, where $p_i\geq 2$. If there exists $i$ such that $1\leq r < \frac{n}{p_i}$ and $r \not\equiv 0 \mod p_i$, then the normalized Laplacian matrices of the complete bipartite graph ${\cal K}_{r,n-r}$ and its complement graph $\overline{{\cal K}_{r,n-r}}$ are entangled in $\C^{p_1}\times \C^{p_2} \times \cdots \times \C^{p_m}$
for all vertex labelings.
\label{thm:bipartite1}
\end{theorem}

\proof  If $A$ is entangled in $\C^{p_1}\times \C^{p_2p_3}$, then it is entangled
in $\C^{p_1}\times \C^{p_2}\times \C^{p_3}$.  So we only need to consider the case
$n = p_1p_2$.  Without loss of generality we assume  $r < p_1$ and $r\not\equiv 0\mod p_2$.
Let the vertices of ${\cal K}_{r,n-r}$ be partitioned into two disjoint sets of vertices $A$ and $B$, with edges from every member of $A$ to every member of $B$ and $|A| = r$.
Since $r\not\equiv 0\mod p_2$, there exists $v_u$ such that
 $(v_u,w_a)$ and $(v_u,w_b)$ are vertices in $A$ and $B$ respectively.  The degree of $(v_u,w_b)$ is $r$ in ${\cal G}$.   Let us look at the degree of $(v_u,w_b)$ in ${\cal G}^{pT}$.  
Consider the vertices $(v_y,w_b)$ for $v_y\neq v_u$.
If $(v_y,w_b)\in A$, then $\{(v_u,w_b),(v_y,w_b)\}$ is an edge of both ${\cal G}$ and 
${\cal G}^{pT}$.
If $(v_y,w_b)\in B$, then $\{(v_u,w_a),(v_y,w_b)\}$ is an edge of ${\cal G}$ and thus 
$\{(v_u,w_b),(v_y,w_a)\}$ is an edge of ${\cal G}^{pT}$.  
Thus we have identified $p_1-1$ edges connected to $(v_u,w_b)$ in ${\cal G}^{pT}$.
Finally $\{(v_u,w_a),(v_u,w_b)\}$ is an edge in both ${\cal G}$ and ${\cal G}^{pT}$.
Thus the degree of $(v_u,w_b)$ in ${\cal G}^{pT}$ is at least $p_1>r$ and thus by Theorem \ref{thm:degree}
the normalized Laplacian matrix is entangled. The part about $\overline{{\cal K}_{r,n-r}}$ follows
from Corollary \ref{cor:degree-comp}.\eop

Note that $\overline{{\cal K}_{r,n-r}} = {\cal K}_r \cup {\cal K}_{n-r}$ is the union of two complete graphs.  Since the normalized Laplacian matrix of a complete graph is separable \cite{braunstein:laplacian:2006}, this means that the union of graphs does not necessarily preserve separability of Laplacian matrices.
If $r\equiv 0 \mod p_2$, then it is easy to find a vertex labeling such that ${\cal K}_{r,n-r}$ is separable in $\C^{p_1}\times \C^{p_2}$.

\begin{corollary}
Let $n=p_1p_2\cdots p_m$, where $p_i\geq 2$. If $1\leq r < \min_i p_i$, then the normalized Laplacian matrices of the graph ${\cal K}_{r,n-r}$ and its complement graph $\overline{{\cal K}_{r,n-r}}$ are entangled in $\C^{p_1}\times \C^{p_2} \times \cdots \times \C^{p_m}$.
\label{cor:bipartite2}
\end{corollary}

\begin{theorem}
Let $n=p_1p_2\cdots p_m > 4$, where $p_i\geq 2$. For all nontrivial complete bipartite graphs
${\cal K}_{r,n-r}$, there exists a vertex labeling such that 
 the normalized Laplacian matrices of  ${\cal K}_{r,n-r}$ and $\overline{{\cal K}_{r,n-r}}$ are entangled in $\C^{p_1}\times \C^{p_2} \times \cdots \times \C^{p_m}$.
\label{thm:bipartite3}
\end{theorem}

\proof As before, we only need to consider the case
$n = p_1p_2$.  Without loss of generality, let us assume that $p_1\leq p_2$ and $r \leq \frac{n}{2}$.
Since ${\cal K}_{r,n-r}$ is nontrivial, $0<r<n$. Let $A$ and $B$ be defined as in Theorem \ref{thm:bipartite1}.  If $r< p_1$, then the result follows from Corollary \ref{cor:bipartite2}.
Suppose $r = p_1$. Assign to the elements of $A$ the labeling $(v_1,w_2), (v_2,w_1), \cdots 
, (v_{p_1},w_1)$. Assign an element of $B$ the labeling $(v_1,w_1)$.  This vertex has degree $r$ in ${\cal G}$.  An edge in ${\cal G}$ from this vertex to each of the vertices of $A$ in will remain an edge in ${\cal G}^{pT}$. Since $p_2\geq 3$, we can assign another vertex in $B$ to
$(v_1,w_3)$.  There is an edge $\{(v_2,w_1),(v_1,w_3)\}$ in ${\cal G}$, so there is
an edge $\{(v_1,w_1),(v_2,w_3)\}$ in ${\cal G}^{pT}$ and $(v_1,w_1)$ has degree at least $r+1$ in ${\cal G}^{pT}$.  By Theorem \ref{thm:degree} the normalized Laplacian matrix is entangled.
Suppose $r> p_1$.  Let $p_1$ elements from $A$ be assigned the labeling $(v_1,w_1),(v_2,w_1), \cdots ,(v_{p_1},w_1)$.  Since
$n-r\geq \frac{n}{2} \geq p_2 > p_2-1$, we can pick $p_2-1$ elements from $B$ and assign them
the labeling $(v_1,w_2), \cdots , (v_1,w_{p_2})$.  Since each element of $A$ is connected to each element of $B$,
$\{(v_i,w_1),(v_1,w_j)\}$ is an edge in ${\cal G}$ for $1\leq i\leq p_1$, $2\leq j\leq p_2$.
Thus $\{(v_1,w_1),(v_i,w_j)\}$ is an edge in ${\cal G}^{pT}$ which means that
$(v_1,w_1)$ is connected to $p_1(p_2-1) = n-p_1$ nodes, i.e. it has degree at least
$n-p_1$ in ${\cal G}^{pT}$.  The vertex $(v_1,w_1)$ is in $A$ so it has degree $n-r < n-p_1$ in ${\cal G}$. Again
the normalized Laplacian matrix is entangled by Theorem \ref{thm:degree}. The part about $\overline{{\cal K}_{r,n-r}}$ follows
from Corollary \ref{cor:degree-comp}.\eop

Ref. \cite{braunstein:laplacian:2006} shows that the bipartite separability of ${\cal K}_{1,n-1}$ and ${\cal K}_n$ do not depend on the vertex labeling. The normalized Laplacian matrix of ${\cal K}_{1,n-1}$  is entangled for all vertex labelings and the normalized Laplacian matrix of ${\cal K}_n$ is separable for all vertex labelings.  It was posed as an open question in Ref. \cite{braunstein:laplacian:2006} whether there are other classes of graphs with this property.  Theorem \ref{thm:bipartite1} and Corollary \ref{cor:bipartite2} list additional classes of graphs whose normalized Laplacian matrices are entangled under any vertex labeling.  

The normalized Laplacian matrix of the complete graph ${\cal K}_n$ is multipartite separable under any vertex labeling (Corollary \ref{cor:complete}).  Are there graphs besides the complete graph whose Laplacian matrix is multipartite separable for all vertex labeling? Theorem \ref{thm:bipartite3} shows that they will not be complete bipartite graphs or their complement graphs.
The results in the following section identify other classes of graphs whose normalized Laplacian matrix is entangled for some vertex labeling.

\subsection{Vertex degree conditions for multipartite entanglement}
For a vertex $v$, let $deg(v)$ denote its vertex degree.
\begin{theorem}
Let $n=p_1p_2\cdots p_m$, where $p_i\geq 2$. Let ${\cal G}$ be a nontrivial graph such that
$\min_w deg(w) < \frac{n}{p_i}-1$ for some $i$.  Then there is a vertex labeling such that the normalized Laplacian matrix of the graph ${\cal G}$ is entangled in $\C^{p_1}\times \C^{p_2} \times \cdots \times \C^{p_m}$.
\label{thm:entangled1}
\end{theorem}
\proof As in Theorem \ref{thm:bipartite1} we only need to consider the case $n = p_1p_2$.
Suppose that $w$ is the vertex with minimal degree $d$ in ${\cal G}$ and without loss of generality, $d < p_2-1$.   We will construct a vertex labeling such that the normalized Laplacian matrix is entangled.  Let $N(w)$ be the set of neighbors of $w$.  By definition,
$|N(w)| = d$.  We assign $w = (v_1,w_1)$, and $(v_1,w_i), 2\leq i\leq d+1$ for the vertices in $N(w)$.  Since $\{(v_1,w_1),(v_1,w_i)\}, 2\leq i\leq d+1$ is an edge
of both ${\cal G}$ and ${\cal G}^{pT}$, $w$ has degree at least $d$ in ${\cal G}^{pT}$.
Finally for a vertex $u\not\in \{w\}\cup N(w)$, $deg(u)\geq d$.  Since the graph is nontrivial, we can find $u$ such that $deg(u) > 0$. There are 2 cases to consider.  In case 1, $u$ is connected to a vertex
$y \not\in \{w\}\cup N(w)$.  We set $u = (v_1,w_{p_2})$ and $y = (v_2,w_1)$.  
This means that $\{(v_1,w_1),(v_2,w_{p_2})\}$ is an edge in ${\cal G}^{pT}$.
In case 2, $u$ is connected to a vertex in $N(w)$, say $(v_1,w_c)$.  We set $u = (v_2,w_1)$.
Then $\{(v_1,w_1),(v_2,w_c)\}$ is an edge in ${\cal G}^{pT}$.  In either case $w$ has degree strictly larger than $d$ in ${\cal G}^{pT}$ and the result follows from Theorem \ref{thm:degree}.\eop

\begin{corollary}
Let $n=p_1p_2\cdots p_m$, where $p_i\geq 2$. Let ${\cal G}$ be a noncomplete graph such that
$\max_w deg(w) > n-\frac{n}{p_i}$ for some $i$.  Then there is a vertex labeling such that the normalized Laplacian matrix of the graph ${\cal G}$ is entangled in $\C^{p_1}\times \C^{p_2} \times \cdots \times \C^{p_m}$.
\label{cor:entangled3}
\end{corollary}
\proof Follows from Theorem \ref{thm:entangled1} and Corollary \ref{cor:degree-comp}.\eop

\begin{theorem}
Let $n=p_1p_2\cdots p_m > 4$, where $p_i\geq 2$. Let ${\cal G}$ be a nontrivial graph such that
$\min_w deg(w) < p_i + \frac{n}{p_i}-2$ for some $i$.  Then there is a vertex labeling such that the normalized Laplacian matrix of the graph ${\cal G}$ is entangled in $\C^{p_1}\times \C^{p_2} \times \cdots \times \C^{p_m}$.
\label{thm:entangled4}
\end{theorem}
\proof We assume without loss of generality that $n = p_1p_2$ and $p_2 > 2$.  
Suppose vertex $w$ has minimal degree $d$.  If $d<p_2-1$ then the result follows from Theorem
\ref{thm:entangled1}.  Therefore we assume that $d\geq p_2-1$.  Assign $w$ to $(v_1,w_1)$.
Let $N(w)$ be the neighbors of $w$ which we partition into two sets $A$ and $B$ of size
$p_2-2$ and $d-p_2+2$ respectively.  Since $d\geq p_2-1$ and $p_2>2$ the sets $A$ and $B$ are both nonempty. We assign vertices in $A$ to
$(v_1,w_2), \cdots , (v_1,w_{p_2-1})$.  We assign vertices in $B$ to
$(v_2,w_1),\cdots , (v_{d-p_2+3},w_1)$.  Note that $d-p_2+3\leq p_1$ by hypothesis.
It is clear that an edge in ${\cal G}$ from $w$ to the elements in $N(w)$ will remain an edge in
${\cal G}^{pT}$.  Thus $w$ has degree at least $d$ in ${\cal G}^{pT}$.
Consider the vertex $(v_2,w_1)$ in $B$.  It has degree $\geq d\geq p_2-1\geq 2$.
This means that it is connected to a vertex $(v_u,w_u) \neq (v_1,w_1)$.
There are 3 cases to consider.  In case 1, $(v_u,w_u)$ is in $A$, i.e. $v_u = v_1$.
Thus there is an edge $\{(v_1,w_u),(v_2,w_1)\}$ in ${\cal G}$ and an
edge $\{(v_1,w_1),(v_2,w_u)\}$ in ${\cal G}^{pT}$.
In case 2, $(v_u,w_u)$ is in $B$.    In this case, switch the assignment with a vertex in $A$ and this reduces it to case 1.
In case 3, $(v_u,w_u) \not\in A\cup B$.  We reassign it to $(v_1,w_{p_2})$.  
Then $\{(v_2,w_1),(v_1,w_{p_2})\}$ is an edge in ${\cal G}$ and  $\{(v_1,w_1),(v_2,w_{p_2})\}$ is an edge in ${\cal G}^{pT}$.
In all cases, the degree of $w$ is strictly larger than $d$ in ${\cal G}^{pT}$ and the result follows from Theorem \ref{thm:degree}.\eop

\begin{corollary}
Let $n=p_1p_2\cdots p_m > 4$, where $p_i\geq 2$. Let ${\cal G}$ be a noncomplete graph such that
$\max_w deg(w) > n - p_i - \frac{n}{p_i}+1$ for some $i$.  Then there is a vertex labeling such that the normalized Laplacian matrix of the graph ${\cal G}$ is entangled in $\C^{p_1}\times \C^{p_2} \times \cdots \times \C^{p_m}$.
\label{cor:entangled5}
\end{corollary}
\proof Follows from Theorem \ref{thm:entangled4} and Corollary \ref{cor:degree-comp}.\eop

\subsection{Bipartite separability in $\C^2\times \C^2$}
\begin{figure}[htbp]
\centerline{\includegraphics[width=3in]{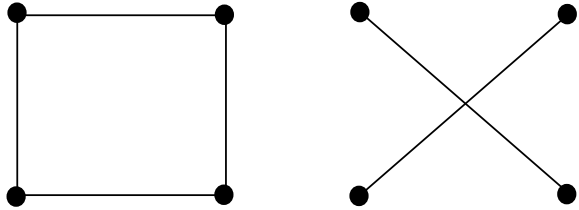}}
\caption{Noncomplete graphs on 4 vertices whose normalized Laplacian matrices are separable regardless of the vertex labeling.}
\label{fig:separable-four}
\end{figure}

In this case $n = 4$ and a simple enumeration (and the fact that the condition in Theorem \ref{thm:degree} is both necessary and sufficient) shows that the complete graph ${\cal K}_4$ and the two graphs in 
Fig. \ref{fig:separable-four} are separable in $\C^2\times \C^2$ for all vertex labelings.  For all other graphs\footnote{We exclude the trivial graph from consideration, since the Laplacian matrix is the zero matrix and has zero trace.} with $4$ vertices, there exists a vertex labeling such that the normalized Laplacian matrix is entangled in $\C^2\times \C^2$.

\subsection{Multipartite separability in $\C^2\times \C^{p_2} \times \cdots \times \C^{p_m}$, $n > 4$}

\begin{theorem}
Let $n = 2p_2p_3\cdots p_m > 4$.
If ${\cal G}$ is a graph that is not complete nor trivial, then
there is a vertex labeling such that the normalized Laplacian matrix is entangled  in $\C^2\times \C^{p_2} \times \cdots \times \C^{p_m}$.
\end{theorem}
\proof  By Theorem \ref{thm:entangled4} such a vertex labeling exists if $\min_w deg(w) \leq 
\frac{n}{2}-1$.  By Corollary \ref{cor:entangled5} such a vertex labeling exists if
$\max_w deg(w) \geq \frac{n}{2}$.  It is clear that one of these two inequalities must be satisfied for any graph.\eop

Computer experiments indicate that for $4 < n \leq 9$, $n$ composite, for all noncomplete graphs there exists a vertex labeling such that the normalized Laplacian matrix is entangled.  It remains to be seen whether this is true for all noncomplete graphs with $n> 4$.

\section{A joint matrix decomposition result}

Let
$I$ denote the identity matrix and $J$ denote the matrix of all $1$'s.    
The $i$-th unit vector is denoted $e_i$.

\begin{definition}
For a complex matrix $A$, let $|A|$ denote the real nonnegative matrix $B$ such that $B_{ij} = |A_{ij}|$.
Let $|A|_{*}$ be the matrix $B$ such that $B_{ii} = A_{ii}$ and $B_{ij} = |A_{ij}|$ for $i\neq j$.
\end{definition}

\begin{theorem}
Let $D$ be a diagonal real matrix and $A$ Hermitian.  If $D-A$ is row diagonally dominant, then
there exists $\lambda_i$, $\mu_i$ real numbers and $v_i$ $n$-vectors such that
\begin{eqnarray} D &=& \sum_i \mu_i v_i v_i^T\nonumber \\
A & = & \sum_i \lambda_i v_i v_i^T  \label{eqn:cond1}\\
\mu_i &\geq & \lambda_i \nonumber
\end{eqnarray} \label{thm:decomp}
\end{theorem}
\proof To prove that the pair $(D,A)$ satisfy the conditions in Eqs. (\ref{eqn:cond1}), we decompose $A$ and $D$ into $A = A_1+A_2$ and $D=D_1+D_2$ such that the pairs $(D_1,A_1)$ and $(D_2,A_2)$ each satisfies the conditions in Eqs. (\ref{eqn:cond1}).  
Let $r_1$,
$r_2$, $\cdots$, $r_n$ be the row sums of $|A|_*$.
Let $A_1 = A - A_2$ where 
\begin{equation}
A_2 = \left(
\begin{array}{ccccc}
0 & & & & \\ &r_2-r_1 & & &\\&& r_3-r_1 &&\\&&&\ddots &\\& & & &r_n-r_1
\end{array}\right)
\end{equation}
It is clear that all row sums of $|A_1|_*$ is equal to $r_1$.  
Let $v_i$ be the normalized eigenvalues of $A_1$ with
eigenvalues $\lambda_i$. 
Thus $A_1 = \sum_i \lambda_i v_1v_1^T$.  Let $D_1 = r_1 I = r_1 \sum_i v_iv_1^T$. By Gershgorin's circle criterion,  $\lambda_i\leq r_1$, so $(D_1,A_1)$ satisfies the conditons in
Eqs. (\ref{eqn:cond1}).
Let $D_2 = D-D_1$.  The row diagonally dominant condition of $D-A$ ensures that $D_2$ is a diagonal matrix such that $(D_2)_{ii}\geq (A_2)_{ii}$.  Since $D_2$ and $A_2$ can be both
expressed as $D_2 = \sum_i (D_2)_{ii}e_ie_i^T$ and $A_2 = \sum_i (A_2)_{ii}e_ie_i^T$, the proof is complete.
\eop

\begin{corollary}
Let $D_j$ be a diagonal real matrix and $A_j$ Hermitian.  If $D_j-A_j$ is row diagonally dominant, then
there exists $\lambda_i$, $\mu_i$ real numbers and $v_i$ $n$-vectors such that
\begin{eqnarray} \sum_j D_j &=& \sum_i \mu_i v_i v_i^T\nonumber \\
\sum_j A_j & = & \sum_i \lambda_i v_i v_i^T  \\
\mu_i &\geq & \lambda_i \nonumber
\end{eqnarray}
\end{corollary}
 
\begin{theorem} \label{thm:separable_main}
Let $D_i$ be  diagonal real matrices and $P_i$ Hermitian matrices in $\C^{p_i\times p_i}$.  Suppose $D_i-P_i$ is row diagonally dominant for $1\leq i\leq m$.  If
$A = D_1\otimes \cdots \otimes D_m - P_1\otimes \cdots \otimes P_m$, then
$\frac{1}{tr(A)} A $ is a separable density matrix in $\C^{p_i}\times \cdots \times \C^{p_m}$, provided $tr(A)\neq 0$.
\end{theorem}

\proof By Theorem \ref{thm:decomp} the matrices
$D_1\otimes \cdots \otimes D_m$ and $P_1\otimes \cdots \otimes P_m$ can be decomposed as
$\otimes_j\sum_i \mu_i^j v_i^j {v_i^j}^T$ and $\otimes_j\sum_i \lambda_i^j v_i^j {v_i^j}^T$
where $\mu_i^j\geq \lambda_i^j$.  This means that $A$ can be written as
$A = \otimes_j\sum_i \left(\mu_i^j-\lambda_i^j\right) v_i^j {v_i^j}^T$, i.e.
$\frac{1}{tr(A)}A$ is separable.
\eop

\begin{corollary} \label{cor:separable_main}
Let $D_i^j$ be diagonal real matrices and $P_i^j$ Hermitian for $1\leq i\leq m$, $1\leq j\leq k$.  Suppose $D_i^j-P_i^j$ is row diagonally dominant for $1\leq i\leq m$, $1\leq j\leq k$.  If
$A = \sum_{j=1}^k D_1^j\otimes \cdots \otimes D_m^j - P_1^j\otimes \cdots \otimes P_m^j$,
then $\frac{1}{tr(A)} A $ is a separable density matrix in $\C^{p_i}\times \cdots \times \C^{p_m}$, provided $tr(A)\neq 0$.
\end{corollary}

\section{Multipartite separability of graph products}

\begin{definition}
For a graph with adjacency matrix $A$, a matrix of the form $L=D-A$ where $D$ is a diagonal real matrix such that $L$ is row diagonally dominant is called a {\em generalized Laplacian matrix} of the graph. \label{def:laplacian}
\end{definition}

Definition \ref{def:laplacian} is different from the definition in \cite{godsil:alg_graph:2001} in that here we can assume $A$ to be complex matrices and require row diagonal dominance.
Clearly this definition does not define $L$ uniquely since there are many choices for the matrix $D$.  However this will not matter for the results in this section.  

In the rest of this section we assume that the adjacency matrix $A$ of a graph is a real nonnegative matrix.
In this case, a matrix element $A_{ij}>0$ can be considered as an edge from vertex $i$ to vertex $j$ with weight $A_{ij}$.  Without loss of generality, we assume that $0\leq A_{ij}\leq 1$.

\begin{definition}
For a complex matrix $A$, $r(A)$ is the diagonal matrix such that $r(A)_{ii} = \sum_{j} A_{ij}$.
\end{definition}
It is clear that $r(|A|)-A$ is a Laplacian matrix of a graph with adjacency matrix $A$.  The following Lemma is easy to show.
\begin{lemma}
$r(A\otimes B) = r(A)\otimes r(B)$, $r(|A\otimes B|) = r(|A|)\otimes r(|B|)$
\label{lem:row}
\end{lemma}

A {\em graph product} of $\cal G$ and $\cal H$ (denoted as ${\cal G}\diamond {\cal H}$)  is defined as a graph with 
vertices $V({\cal G})\times V({\cal H})$ and edges defined by:

$\{(u,v),(w,y)\}$ is an edge if and only if $Q$ is true.  The relation $Q$ is of the form  $P_1 \vee P_2 \vee \cdots$.
where each $P_i$ is one of $8$ conditions:
\begin{itemize}
\item $R_1: (u \quad\mbox{adj} \quad w) \wedge (v \quad\mbox{adj} \quad y)$
\item $R_2: (u \quad\mbox{adj} \quad w) \wedge (v = y)$
\item $R_3: (u \quad\mbox{adj}\quad w) \wedge (v \quad\neg \mbox{adj} \quad y)$
\item $R_4: (u = w) \wedge (v \quad\mbox{adj} \quad y)$
\item $R_5: (u = w) \wedge (v \quad\neg \mbox{adj} \quad y)$
\item $R_6: (u \quad\neg \mbox{adj} \quad w) \wedge (v \quad\mbox{adj}\quad y)$
\item $R_7: (u \quad\neg \mbox{adj} \quad w) \wedge (v = y)$
\item $R_8: (u \quad\neg \mbox{adj} \quad w) \wedge (v \quad\neg \mbox{adj} \quad y)$
\end{itemize}
Thus there are $2^8 = 256$ different types of graph products.
Table \ref{tbl:products} lists some possibilities for $Q$ and the names associated to the corresponding graph product \cite{bray:graph_products}.

\begin{table}[htbp]
\caption{Commonly used graph products.}
\centering
\begin{tabular}{|c|c|}
\hline
Condition $Q$ & name(s) \\
\hline
$R_1$ & tensor product, categorical product,\\
& direct product, cardinal product \\
$R_1 \vee R_2 \vee R_4$ & strong product \\
$R_2 \vee R_4$ & Cartesian product \\
$R_1 \vee R_2 \vee R_3 \vee R_4$ & lexicographical product \\
\hline
\end{tabular}
\label{tbl:products}
\end{table}

Let the adjacency matrices of $\cal G$ and $\cal H$ be $G$ and $H$ respectively.
It is easy to show that the adjacency matrix of a graph product ${\cal G}\diamond{\cal H}$ is $\sum_i T_i$ where
to each condition $P_i$ in $Q$ corresponds a matrix according to Table \ref{tbl:graphprod}.

\begin{table}[htbp]
\caption{Matrices $T_i$ corresponding to each of the $8$ conditions $R_i$.}

\centering
\begin{tabular}{|c|c|}
\hline
Condition & $T_i$ \\
\hline
$(u \quad\mbox{adj} \quad w) \wedge (v \quad\mbox{adj} \quad y)$ & $G\otimes H$\\
$(u \quad\mbox{adj} \quad w) \wedge (v = y)$ & $G\otimes I$ \\
$(u \quad\mbox{adj} \quad w) \wedge (v \quad\neg \mbox{adj} \quad y)$ & $G \otimes (J-I-H)$ \\
$(u = w) \wedge (v \quad\mbox{adj}\quad y)$ & $I\otimes H$ \\
$(u = w) \wedge (v \quad\neg \mbox{adj} \quad y)$ & $I\otimes (J-I-H)$ \\
$(u \quad\neg \mbox{adj} \quad w) \wedge (v \quad\mbox{adj}\quad y)$ & $(J-I-G)\otimes H$ \\
$(u \quad\neg \mbox{adj} \quad w) \wedge (v = y)$ & $ (J-I-G)\otimes I$ \\
$(u \quad\neg \mbox{adj} \quad w) \wedge (v \quad\neg \mbox{adj} \quad y)$ & $(J-I-G)\otimes (J-I-H)$ \\
\hline
\end{tabular}
\label{tbl:graphprod}

\end{table}

\begin{theorem}
The complement of a graph product is another graph product.
\end{theorem}
\proof We'll prove this for a graph product of two graphs, as the general case is similar. 
The adjacency matrix of ${\cal G} \diamond {\cal H}$ is $\sum_i T_i$.  Its complement graph has adjacency matrix $J - I - \sum_i T_i = J \otimes J - I\otimes I - \sum_i T_i$.
 Let $c(P) = J-I-P$.  The matrix $J$ can be decomposed as $J = c(P) + I + P$ and
 $J \otimes J - I\otimes I= (c(G) + I + G)\otimes (c(H) + I + H) - I\otimes I$ is exactly the sum of the 8 possible $T_i$'s in Table \ref{tbl:graphprod}.  This means that 
the adjacency matrix of the complement graph is also of the form $\sum_i T_i$ and the proof is complete.\eop

The following theorem shows that the normalized Laplacian matrix of a graph product is multipartite separable.
\begin{theorem}
For a set of graphs ${\cal P}_i$, if $A$ is a Laplacian matrix of ${\cal P}_1 \diamond {\cal P}_2 \diamond \cdots \diamond {\cal P}_m$, where $\diamond$ is a graph product and $tr(A)\neq 0$, then
$\frac{1}{tr(A)}A$ is a separable density matrix in $\C^{p_1}\times \dots \times 
\C^{p_m}$, where $p_i$ is the order of ${\cal P}_i$. \label{thm:separable_product}
\end{theorem}
\proof Let $B$ be the adjacency matrix of ${\cal P}_1 \diamond {\cal P}_2 \diamond \cdots \diamond {\cal P}_m$.  By Lemma \ref{lem:row}, the diagonal matrix of the row sums of $|B|$ can be written as  $\sum_{j} \otimes_i D_i^j$  and $A$ can be written as $\sum_{j} \otimes_i D_i^j - \sum_j \otimes T_i^j$.  The theorem is then a direct consequence of Corollary \ref{cor:separable_main}.\eop

The special cases of tensor products ($Q=R_1$) and $m=2$ (bipartite separability), $m=3$ (tripartite separability) were proven in \cite{braunstein:laplacian:2006} and \cite{wang:tripartite:2007} respectively.  The same proof shows that Theorem \ref{thm:separable_product} is true for the case of 
tensor products even when the adjacency matrices are complex matrices.

\begin{corollary}
Let $A = nI-J$ be the Laplacian matrix of the complete graph ${\cal K}_n$, and $n = p_1p_2\cdots p_m$.  Then
$\frac{1}{tr(A)}A$ is a separable density matrix in $\C^{p_1}\times \dots \times 
\C^{p_m}$. \label{cor:complete}
\end{corollary}
\proof This follows from Theorem \ref{thm:separable_product} and the fact that the complete graph ${\cal K}_{pq}$ is the strong product of ${\cal K}_p$ and ${\cal K}_q$.
\eop

Corollary \ref{cor:complete} for the cases of $m=2$ and $m=3$ were proven in \cite{braunstein:laplacian:2006} and \cite{wang:tripartite:2007} respectively.

\section{Conclusions}
We continue the study of normalized Laplacian matrices of graphs as density matrices and analyze their entanglement properties.  In particular, we identify graphs whose normalized Laplacian matrices are entangled for every vertex labeling or whose Laplacian matrices are entangled for some vertex labeling.  Furthermore, we show that normalized Laplacian matrices of  graph products are multipartite separable.  

\bibliographystyle{IEEEtran}
\bibliography{quantum,graph_theory,algebraic_graph}
\end{document}